\documentstyle[12pt]{article}
\newcommand{\be}{\begin{equation}}
\newcommand{\ee}{\end{equation}}
\newcommand{\bea}{\begin{eqnarray}}
\newcommand{\eea}{\end{eqnarray}}

\newcommand{\rme}{\mbox{e}}
\newcommand{\rmi}{{\rm i}}
\newcommand{\cZ}{{\cal Z}}
\newcommand{\cE}{{\cal E}}
\newcommand{\cN}{{\cal N}}

\begin{document}
\begin{titlepage}

\begin{flushright}
Oslo SHS-96-2 \\
hep-th/9602171 \\
April 23, 1996
\end{flushright}
\vspace*{\fill}
\centerline{\LARGE\bf
The third virial coefficient of anyons revisited
}

\renewcommand{\thefootnote}{\fnsymbol{footnote}}
\vspace{2cm}
\centerline{\large Stefan Mashkevich$^{a,c,}$\footnote{
mash@phys.unit.no, janm@phys.unit.no, kolausen@phys.unit.no}}
\vspace{0.3cm}
\centerline{\large Jan Myrheim and K{\aa}re Olaussen$^{b,c,}$\footnotemark[1]}
\vspace{1.2cm}
\centerline{\large ${}^a$ \it Institute for Theoretical Physics,
252143 Kiev, Ukraine}
\vspace{0.3cm}
\centerline{\large ${}^b$ \it Department of Physics,
NTNU, N--7034 Trondheim, Norway}
\vspace{0.3cm}
\centerline{\large ${}^c$ \it Centre for Advanced Study,
Drammensveien 78} 
\centerline{\large \it N--0271 Oslo, Norway}
\vspace{3cm}

\centerline{\bf Abstract}
\vspace{0.5cm}
We use the method of solving the three-anyon problem developed in
our earlier publication 
to evaluate numerically the third virial coefficient of free anyons.
In order to improve precision, we explicitly correct for
truncation effects.
The present calculation is about three orders of magnitude more
precise than the previous Monte Carlo calculation
and indicates the presence of a term 
$a \sin^4 \pi\nu$ with a very small coefficient
$a \simeq -1.65\times 10^{-5}$.
\vspace{0.4cm}
\vspace*{\fill}
\end{titlepage}

\section{Introduction}

The three-anyon problem, by now, has a rather long history
(see \cite{mmor1} and references therein). It is quite
interesting as an example of a two-dimensional
three-body problem which is apparently integrable,
and besides that, it provides a means of determining
the third virial coefficient of anyons.
The virial expansion gives the pressure $P$ in
terms of the particle density $n$,
\be
P\beta = n[ 1 + A_2(\lambda^2n) + A_3(\lambda^2n)^2
+ \ldots \,]
\ee
($\beta$ is the inverse temperature, $\lambda$ is the thermal
wavelength) and shows the deviation of the
equation of state from that of a classical ideal gas.
Here $A_2$, $A_3$ etc.\ are the (dimensionless) virial coefficients.
For anyons, they depend on the statistics parameter $\nu$.

The second virial coefficient of anyons, which can be found from
the solution of the two-anyon problem, is known exactly \cite{2nd},
\be
A_2(\nu)=\frac{1}{4} - \frac{(1-\nu)^2}{2} \; , \qquad 0\le\nu\le2
\ee
($\nu=0$ and $2$ correspond to bosons, $\nu=1$ to fermions).

The problem of finding the third virial coefficient 
amounts, in one way or another, to solving the three-anyon
problem and is therefore non-trivial.
It is convenient to consider the difference
\be
\Delta A_3(\nu) = A_3(\nu) - A_3(0)\; ,
\label{delta}
\ee
where
\be
A_3(0) = A_3(1) = \frac{1}{36}
\ee
is the third virial coefficient of bosons and fermions.
On the basis of numerical calculations,
it has been conjectured \cite{mo,mthes} that $\Delta A_3(\nu)$
obeys a remarkably simple formula,
\be
\Delta A_3(\nu) = \frac{1}{12\,\pi^2}\,\sin^2 \pi\nu\; .
\label{conj}
\ee
It is consistent with the second-order perturbative result
$\Delta A_3(\nu) \simeq \nu^2/12$ \cite{pert}, with the exact
supersymmetry property $\Delta A_3(\nu) = \Delta A_3(1-\nu)$
\cite{ss} (which means that it is enough to consider $\nu\le0.5$), 
and with the numerical Monte Carlo results \cite{mo,mthes} 
which are accurate to about 5\% for $\nu$ close to $0.5$\,.
An independent check of (\ref{conj}) by means of a numerical
calculation of the energy levels has been carried out \cite{numold,num}, 
confirming it for $\nu\le0.25$ with an accuracy of about 1\%
(the accuracy of the method being worse beyond this interval).

This letter is devoted to a calculation of the third virial
coefficient by finding the levels numerically using the method
developed in our earlier paper \cite{mmor1}.
With the inaccuracy in $\Delta A_3(\nu)$ estimated to be of
the order of $0.005\%$ for $\nu$ close to $0.5$, our results fit
well to the modified formula
\be
\Delta A_3(\nu) = \frac{1}{12\,\pi^2}\,\sin^2 \pi\nu
+ a \sin^4 \pi\nu
\label{ourres}
\ee
with a small coefficient $a = -(1.652 \pm 0.012)\times 10^{-5}$.
This means that the task of calculating
$\Delta A_3(\nu)$ analytically appears more difficult than
if (\ref{conj}) were exact.

\section{The method}

Recall some basic equations concerning the third virial coefficient.
It is most convenient to use the harmonic oscillator regularization,
whence
\be
A_3(\nu) = -2\lim_{x \rightarrow 0} \left[ 3\tilde{Z}_3
- 3Z_2 + Z_1^2 \right] + 4\left[A_2(\nu)\right]^2\;,
\label{2.1}
\ee
where
\be
Z_1 = \frac{\rme^x}{(\rme^x-1)^2}\; , \qquad
Z_2 = \frac{\rme^{2x}(\rme^{\nu x} + \rme^{-\nu x})}{(\rme^{2x}-1)^2}
\label{2.2}
\ee
are the one- and two-particle partition functions, respectively,
$\tilde{Z}_3 = Z_3/Z_1$
is the three-particle relative motion
partition function, and $x=\hbar\beta\omega$
($\omega$ is the harmonic frequency).
The three-anyon spectrum consists of two types of states,
interpolating linearly or
nonlinearly between the bosonic and fermionic points,
thus $\tilde{Z}_3$ can be split into two parts,
\be
\tilde{Z}_3 = \tilde{Z}_3^{\rm L} + \tilde{Z}_3^{\rm NL}\;,
\ee
where the contribution of the linear states is known exactly, 
\be
\tilde{Z}_3^{\rm L} = \frac{\rme^{5x}(\rme^{3\nu x} + \rme^{-3\nu x})}
{(\rme^{2x}-1)^2(\rme^{3x}-1)^2}
\ee
and the contribution of the nonlinear ones, $\tilde{Z}_3^{\rm NL}$,
has to be evaluated. Since there is the so-called tower structure
of the spectrum \cite{ss,tower}, all the states coming in towers with
the same angular momentum and with energy spacing $2\hbar\omega$, it
can be expressed as
\be
\tilde{Z}_3^{\rm NL} = \frac{\cZ}{1-\rme^{-2x}}\;,
\label{2.5}
\ee
where $\cZ\equiv\cZ(\nu,x)$ is the contribution of the bottom states
(lowest states in the towers).
Substituting (\ref{2.2})--(\ref{2.5}) into (\ref{2.1})
and then into (\ref{delta}) leads to the expression (cf.~\cite{num})
\be
\Delta A_3(\nu) = 
 -6 \lim_{x\to0} \frac{\cZ(\nu,x) - \cZ(0,x)}{1-\rme^{-2x}}\;.
\label{2.6}
\ee

We compute $\cZ(\nu,x)$ according to its definition,
\be
\cZ(\nu,x) = \sum_s \rme^{-xE_s(\nu)}\;,
\label{2.7}
\ee
where $s$ labels the nonlinear bottom states, and $E_s(\nu)$ are their
dimensionless energies, obtained by setting $\hbar\omega=1$.
These energies are found numerically making
use of the method of \cite{mmor1}, the essence of which is as
follows. The complex particle coordinates $z_j = (x_j+\rmi y_j)/\sqrt 2, \:
j=1,2,3$ are transformed into $Z,u,v$ according to
$$
\left( \begin{array}{c} Z \\ u \\ v \end{array} \right) = \frac{1}{\sqrt 3}
\left( \begin{array}{ccc} 1 & 1 & 1 \\ 1 & \eta & \eta^2 \\ 
1 & \eta^2 & \eta \end{array} \right)
\left( \begin{array}{c} z_1 \\ z_2 \\ z_3 \end{array} \right),
$$
where $\eta = \exp(2\pi\rmi/3)$. The center-of-mass coordinate
$Z$ separates from the relative coordinates $u,v$,
which are further rewritten as
$$
u = \frac{rq\,\rme^{\rmi(\varphi+\xi/6)}}{\sqrt{2(1+q^2)}}\;,\qquad
v=\frac{r\,\rme^{\rmi(\varphi-\xi/6)}}{\sqrt{2(1+q^2)}}\;.
$$
{\sloppy
A common eigenstate of the relative Hamiltonian and
angular momentum satisfying the cyclic interchange
anyonic boundary condition, is
$\psi(r,q,\varphi,\xi)=r^{\mu} P_n(r^2) \exp(-r^2/2)
\,\Omega(q,\varphi,\xi)$,
where $P_n$ is a certain polynomial of degree $n$ and the angular
part is

}
\be
\Omega(q,\varphi,\xi) = \rme^{\rmi L\varphi}
\sum_{m=-\infty}^{\infty} \gamma_m g_m(q) \rme^{-\rmi(m+\nu'/2)\xi} 
\label{angpart}
\ee
with
\be
g_m(q)=q^{|j|}(1+q^2)^{-\mu/2}{}_2F_1\left(
\frac{|j|+|k|}{2}-\frac{\mu}{2}\, ,
\frac{|j|-|k|}{2}-\frac{\mu}{2}\: ;
1+|j|\: ;-q^2 \right),
\label{gmq}
\ee
$$
j={L\over 2}-3\left(m+{\nu'\over 2}\right),\qquad
k={L\over 2}+3\left(m+{\nu'\over 2}\right)\;.
$$
We have put $\hbar$, $\omega$ and the particle mass equal to 1.
The angular momentum is $L=L_0+3\nu$, where $L_0$ must be an
integer, $\nu' = \nu + {L_0}\bmod{2}$, and the energy eigenvalue is
\be
E = \mu + 2n + 2\;.
\label{E}
\ee
The radial quantum number $n$ is responsible for the tower structure,
and the bottom states have $n=0$. In the angular part of the wave
function, the coefficients $\gamma_m$ have to be chosen in such a
way that it would satisfy the two-particle exchange
anyonic boundary conditions, which have the form \cite{mmor1}
\begin{eqnarray}
\sum_{m=-\infty}^{\infty}
\gamma_m g_m(1) \,\rme^{\rmi m\xi} & = & \rme^{\rmi(\nu\pi-\nu'\xi)}
\sum_{m=-\infty}^{\infty}
\gamma_m g_m(1) \,\rme^{-\rmi m\xi}\;, \label{bc1} \\
\sum_{m=-\infty}^{\infty}
\gamma_m g'_m(1) \,\rme^{\rmi m\xi} & = & -\rme^{\rmi(\nu\pi-\nu'\xi)}
\sum_{m=-\infty}^{\infty}
\gamma_m g'_m(1) \,\rme^{-\rmi m\xi} \label{bc2}
\end{eqnarray}
[$\;g'_m(q) \equiv dg_m(q)/dq\;$].

To use these boundary conditions, introduce a linear operator $A$ such that
\be
[A\phi](\xi)=\rme^{\rmi(\nu\pi-\nu'\xi)}\phi(2\pi-\xi)\;.
\label{A}
\ee
We take the function $\phi(\xi)$ to be defined for $\xi\in[0,2\pi]$ and
we represent it
by the vector of its Fourier components $\{\phi_m\}$,
so that $\phi(\xi) = \sum_{m=-\infty}^{\infty} \phi_m
\rme^{\rmi m\xi}$.
The matrix elements of $A$ in this representation are 
\be
A_{mn} = \frac{\sin\pi\nu}{\pi(m+n+\nu')}\;.
\label{matel}
\ee
Next introduce two diagonal matrices
\be
G_{mn} = g_m(1)\,\delta_{mn}\;, \qquad
G'_{mn} = g'_m(1)\,\delta_{mn}\;.
\label{G}
\ee
The two boundary conditions (\ref{bc1})--(\ref{bc2}) then take the form
\begin{equation}
(I-A)G\gamma = 0\;, \qquad (I+A)G'\gamma = 0\;,
\label{two}
\end{equation}
where $\gamma$ is the vector $\{\gamma_m\}$ of the unknown
coefficients in the wave function (\ref{angpart}).
Now, since $A$ is a real symmetric matrix and
$A^2=I$, the vectors on the left-hand sides of
the two equations in (\ref{two}) are orthogonal.
This makes it possible to replace the two equations by one,
\be
[G + G' + A(G'-G)]\gamma = 0 \; .
\ee
A nontrivial solution for $\gamma$ exists if and only if
the determinant, which depends on $\mu$ through
(\ref{G}) and (\ref{gmq}), vanishes.

Thus, in order to find the energy eigenvalues,
one fixes $\nu$, chooses a sector of bosonic angular momentum
$L_0$ and scans an interval of $\mu$ in order to find the zeros
$\mu_s(\nu)$ of the determinant. After repeating the procedure for
different $L_0$ and dropping the known linear states, one
gets the eigenvalues $E_s(\nu) = \mu_s(\nu)+2$,
to be substituted into (\ref{2.7}). A good consistency check of
our numerical results is that the number of states found
agrees with the exactly known multiplicities \cite{multi}.

\section{Truncation effects}

There are two sources of error in our calculation 
even at infinite numerical precision. The first one is only a finite
number of Fourier coefficients being considered,
the second one is only states with energies less than
a certain finite $\cE$ being included (``energy cutoff'').
The two effects are independent, and in order
to improve precision we will explicitly take them into account.

We truncate the Fourier series at $|m|=\cN$, i.e., consider $(2\cN+1)$
Fourier components $\{ \gamma_{-\cN},\gamma_{-\cN+1},\ldots,\gamma_{\cN} \}$.
To maintain consistency, one has then to restrict oneself to a
discrete set $\{ \xi_{-\cN},\xi_{-\cN+1},\ldots,\xi_{\cN} \}$
instead of a continuous variable $\xi$, so that any set
of function values $\gamma(\xi_n)$ can be represented
by the Fourier components exactly.
Choosing $\xi_n = \pi + 2\pi n/(2\cN+1)$ and demanding (\ref{A}) to
hold at the points $\xi_n$, one gets for the matrix elements of $A$
\be
A_{mn} = \frac{\sin\pi\nu}{(2\cN+1)\sin[\pi(m+n+\nu')/(2\cN+1)]}\;,
\ee
with the correct $\cN\to\infty$ limit (\ref{matel}).

By analyzing the data $E_s(\nu,\cN)$ obtained at different $\cN$,
we find that they are rather well described by an empirical formula
\be
E_s(\nu,\cN) = E_s(\nu) + \frac{{\cal A}_0}{\cN^{2\nu}}
+ \frac{{\cal A}_1}{\cN^{2\nu+1}} + \ldots \; ,
\label{extrapn}
\ee
which allows to extrapolate to $\cN=\infty$.

Now assume that the levels are found exactly and consider the effect of
the energy cutoff. In practice one can only find the quantity
\be
\cZ(\nu,x,\cE) = \sum_{E_s(\nu) < {\cal E}} \rme^{-xE_s(\nu)}
\label{4}
\ee
for a finite $\cE$, so that (\ref{2.6}) should be rewritten as
\be
\Delta A_3(\nu) = \lim_{x\to0} \lim_{{\cal E}\to\infty}
\Delta A_3(\nu,x,\cE)\;,
\label{4bis}
\ee
where
\be
\Delta A_3(\nu,x,\cE)
= -6 \,\frac{\cZ(\nu,x,\cE) - \cZ(0,x,\cE)}{1-\rme^{-2x}}\;.
\label{3}
\ee
Note that the order of limits cannot be interchanged.
Of course, $\cZ(0,x)$ is known exactly, but truncating it at the
same $\cE$ removes a spurious divergence.

We now argue that
\be
\Delta A_3(\nu,x,\cE) = \Delta A_3(\nu,x)
+ \rme^{-\cE x}f(\nu,x,\cE)\;,
\label{c1}
\ee
where $f(\nu,x,\cE)$ grows (or diminishes) with $x$ not faster than
$\rme^{kx}$ with $|k|\sim1$ and has a weak (polynomial type)
dependence on $\cE$. 
Indeed, the number of nonlinear bottom states grows with energy $E$
like $E^2/9$ \cite{multi}, and, respectively,
$$
\cZ(0,x,\cE) = 
\frac{-18\,\rme^{3x} + \rme^{-\cE x}
\left[(\cE^2 - 3\cE) + (-2\cE^2 + 18)\rme^{3x}
+ (\cE^2 + 3\cE)\rme^{6x}\right]}
{9\,(1-\rme^x)(1-\rme^{3x})^2}
$$
(up to terms proportional to $\cE\bmod3$). Since when $\nu$ changes,
energies change by values of the order of unity
but the number of states is conserved, $\cZ(\nu,x,\cE)$
should still have the form $\cZ(\nu,x) + \rme^{-\cE x}{\cal F}(\nu,x,\cE)$
with $\cal F$ changing not faster than $\cE^2$ and not faster than
$\rme^{kx}$. This leads to the statement expressed by eq.~(\ref{c1}).
In fact, $\cal F$ will be singular at $x\to0$ for a finite $\cE$,
but all the singularities cancel out when doing the subtraction in (\ref{3}).

On the other hand, at small $x$ there has to be an expansion
\be
\Delta A_3(\nu,x) \equiv \lim_{\cE\to\infty}
\Delta A_3(\nu,x,\cE) = a_0 + a_2x^2 + a_4x^4 + \ldots \;.
\label{c2}
\ee
For a discussion of why we include only even powers,
see the next section.
Substituting (\ref{c2}) into (\ref{c1}) and making a power
series expansion for $f(\nu,x,\cE)$, one has that
\be
\Delta A_3(\nu,x,\cE) = a_0 + a_2x^2 + a_4x^4 + \ldots
+ \rme^{-\cE x}(b_0 + b_1x + b_2x^2 + \ldots\:)\;.
\label{da3}
\ee

Remember that for any finite $\cE$,
$\Delta A_3(\nu,x,\cE)$ is {\it finite}\/ at $x=0$;
this is in contrast to ref.~\cite{num} where it was
treated as being infinite.
In fact, expanding the numerator and denominator of
(\ref{3}) in powers of $x$, one gets
\begin{eqnarray}
\Delta A_3(\nu,x,\cE)
& = & -6\,\frac{\sum_{n=1}^\infty (-1)^{n+1} S_nx^n/n!}
{\sum_{n=1}^\infty (-1)^{n+1}(2x)^n/n!}
\nonumber \\
& = & 3S_1 + \left( 3S_1-\frac{3}{2}S_2 \right)x
+ \left( S_1-\frac{3}{2}S_2+\frac{1}{2}S_3 \right)x^2 + \ldots\:,
\label{smx}
\end{eqnarray}
where
\be
S_n \equiv S_n(\nu,\cE) = \sum_{E_s(\nu) < {\cal E}}
\left[ E_s^n(\nu) - E_s^n(0)\right]
\label{sn}
\ee
is finite for any $n$. 
The ansatz (\ref{da3}) must be made consistent with (\ref{smx}).
Matching coefficients at equal powers of $x$ after expanding
the exponential in (\ref{da3}) 
leads to the following expressions for the coefficients $b_n$:
\begin{eqnarray*}
b_0 & = & 3S_1 - a_0\;, \\
b_1 & = & (3+3\cE)S_1 - \frac{3}{2}S_2 - {\cal E}a_0\;, \\
b_2 & = & \bigg(1 + 3\cE + \frac{3\cE^2}{2}\bigg)S_1 
- \bigg(\frac{3}{2} + \frac{3\cE}{2}\bigg)S_2 + \frac{1}{2}S_3
- \frac{\cE^2}{2}a_0 - a_2\;, \\
& \ldots & 
\end{eqnarray*}
This leaves only $a_0,a_2,\ldots$ as unknown coefficients.

We now retain only a finite number of coefficients $a_0,a_2,\ldots,a_{2N}$
and $b_0,b_1,\ldots,b_{2N}$ in (\ref{da3}). The truncated equation can
be rewritten as
\be
\Delta A_3^-(\nu,x,\cE) = \sum_{n=0}^N a_{2n}f_{2n}(x,\cE)\;,
\label{2fit}
\ee
with all the unknown coefficients on the right-hand side and all the
terms depending on the energies $E_s(\nu)$ on the left-hand side.
The quantity on the left-hand side is
\be
\Delta A_3^-(\nu,x,\cE) \equiv \Delta A_3(\nu,x,\cE)
- \sum_{n=0}^{2N} c_nx^n\;
\label{minus}
\ee
with
\begin{eqnarray*}
c_0 & = & 3S_1 \;, \\
c_1 & = & (3+3\cE)S_1 - \frac{3}{2}S_2 \;, \\
& \ldots & 
\end{eqnarray*}
and the functions on the right-hand side depend only on $x$ and $\cE$,
\be
f_{2n}(x,\cE) = x^{2n}\left[1 - \rme^{-\cE x}
\sum_{m=0}^{2N-2n}(\cE x)^m/m!\right].
\ee

From eq.~(\ref{2fit}) with
$\Delta A_3^-(\nu,x,\cE)$ calculated numerically,
one obtains the coefficients $a_{2n}$ through a least square fit
over a certain $x$ interval, and by virtue of
(\ref{4bis}) and (\ref{da3}) one has $\Delta A_3(\nu) = a_0$.
Note that at $x=0$, each of the functions $f_{2n}(x,\cE)$ vanishes
together with its first $2N-1$ derivatives with respect to $x$.
Indeed, by the subtraction in (\ref{minus}) one has already
taken into account the small $x$ behavior of $\Delta A_3(\nu,x,\cE)$,
so that $\Delta A_3^-(\nu,x,\cE)$ has to vanish at $x=0$
up to order $2N-1$. But fixing a positive $x$ and taking
$\cE\to\infty$, as in (\ref{4bis}), results in $f_{2n}(x,\cE)\to x^{2n}$,
meaning that $\Delta A_3^-(\nu,x,\cE)$ tends to the ${\cal E}=\infty$
form (\ref{c2}).

\section{Results and discussion}

We take 6 different values $\cN=10,20,40,80,160,320$ and the energy
cutoff up to $\cE=35$.
The $\cN$ convergence is better for $\nu$
closer to 1---which is natural, because then the wave functions
$\Omega(1,\phi,\xi)$ are less singular at $\xi=0,2\pi$---and
therefore we do all the calculations
on the (semion--fermion) interval $\nu\in[0.5,1]$; that is, the
values listed for $\nu$ are actually obtained at $1-\nu$.
Taking six different values of $\cN$ allows to include five
correction terms in (\ref{extrapn}).

The procedure of extracting $\Delta A_3(\nu)$ is illustrated in Fig.\ 1,
for the case of semions, $\nu=0.5$. 
(In all the figures, the data for $\Delta A_3$ are multiplied
by $10^4$, for convenience.)
The $x$ expansion was truncated at $2N=8$. As $\cE$ is increased,
$\Delta A_3^-(\nu,x,\cE)$ approaches the $\cE=\infty$ curve at any
finite $x$, but always vanishes at $x=0$. We emphasize that, unlike in
ref.~\cite{num}, we do not need an $x$ cutoff on the left; the fit
is performed on an interval $x\in[0,x_{\rm max}]$, and the points
at small $x$, where the initial data have nothing in common with
the $\cE=\infty$ curve, are fully accounted for.

We take different $\cal E$ and $x_{\rm max}$ and observe how the
extracted $\Delta A_3(\nu)$ depends on those. Ideally, there should
be no dependence at all, thus the dependence left over provides an
estimate of the error of the calculation. Figure 2 shows this dependence
for semions. For $x_{\rm max}$ too small, the results of the fitting
become senseless, since one tries to extract too much information from
too narrow a range of data, while for $x_{\rm max}$ too large, the
effect of truncation of the $x$ expansion becomes significant.
It is seen, however, that there exists an interval of $x_{\rm max}$
where $\Delta A_3(\nu)$ is practically constant. Shown in the figure
is an ``error bar'' that we take for the result. Increasing or
decreasing $2N$ leads, in accordance with the aforesaid, to the
``good'' interval of $x_{\rm max}$ moving to the right or to the
left, respectively, but the result does not change significantly
for $2N$ changing down to 4 or up to 10.
On the other hand, the $\cE$ and $x_{\rm max}$ dependence
becomes considerably stronger when the data for $\cN = 160$ and/or $320$
are not used.
Thus, the main source of error appears to be in the inaccuracy
of individual levels.

We want to emphasize that our error analysis by varying the parameters
$\cal N$, $\cal E$ and $x_{\rm max}$ is at the same time a justification
of the methods we use for obtaining the limits as ${\cal N}\to\infty$,
${\cal E}\to\infty$ and $x\to0$. We argue that the result
for $\Delta A_3(\nu)$ is correct to the degree to which it is independent
of $\cal E$, $x_{\rm max}$, and the maximal value of $\cal N$. As for the
limit $x\to0$ in eq.~(\ref{4bis}), if eq.~(\ref{da3}) did not represent
the true $x$ dependence, we would expect the fitted values of the
coefficients $a_{2n}$ in eq.~(\ref{2fit}) to depend on the interval
$0<x<x_{\rm max}$ where the fit is done. Since there exists a finite
range of $x_{\rm max}$ where they depend very little on $x_{\rm max}$,
we believe that these are close to the true values.

The omission of odd powers of $x$ in eq.~(\ref{c2})
(like in ref.~\cite{num}) is only empirically justified,
apart from a general argument \cite{ll} to the
effect that the linear term must be absent, as this
expansion results from an $\hbar$ expansion of the partition
functions (we thank Avinash Khare for pointing this out).
As a second check, we tried to include the terms $x^3, x^5, \ldots \;$.
This changed the values of $\Delta A_3(\nu)$ by amounts
typically about twice our quoted errors, while the coefficients of these
terms were one to two orders smaller than those of the adjacent
even order terms, and were minimal for values of $x_{\rm max}$ where
$\Delta A_3(\nu)$ depends most weakly on $x_{\rm max}$.
This is a clear indication that these terms are really absent,
so that the fit where they are omitted should be more precise.

Table 1 and Fig.~3 contains the main result of this paper.
Shown there is the
difference $\Delta A_3(\nu) - (1/12\pi^2)\sin^2\pi\nu$
together with the best fit curve of the form $a\sin^4\pi\nu$.
The result is
\be
a = -(1.652 \pm 0.012)\times 10^{-5} = -\frac{1}{(621\pm5)\pi^4} \; .
\ee
The value of $\chi^2$ per degree of freedom is $1.0$.
If we exclude the three points closest to $\nu=0$, it drops to
$0.45$, which we regard as a strong indication that the errors are
overestimated for $\nu$ close to $0.5$. This is in contrast
to ref.~\cite{num}, where the results for $\nu$ close to $0.5$ were
the least accurate.
We have included the value $\nu=(3-\sqrt5)/2 \simeq 0.38$, the
golden ratio, as an example of an irrational number which is
badly approximated by rational numbers. This point falls on the
same smooth curve as all the points at rational values of $\nu$.

\begin{center}
\begin{tabular}{|c|r|r|}
\hline
$\nu$ & Our result & Eq.~(\ref{conj}) \\
\hline
0.05 & 2.063(2)  & 2.066  \\
0.10 & 8.057(3)  & 8.063  \\
0.15 & 17.391(3) & 17.403 \\
0.20 & 29.149(2) & 29.171 \\
0.25 & 42.174(2) & 42.217 \\
0.30 & 55.192(2) & 55.263 \\
0.35 & 66.927(3) & 67.032 \\
0.38 & 73.222(2) & 73.347 \\
0.40 & 76.237(2) & 76.372 \\
0.45 & 82.212(2) & 82.368 \\
0.50 & 84.270(2) & 84.434 \\
\hline
\end{tabular}
\end{center}

\begin{center}
Table 1. The values of $10^4 \Delta A_3(\nu)$.
\end{center}

Shown in Fig.~4 for comparison are the Monte Carlo data obtained by
the method of ref.~\cite{mo} from 330 million paths at
$x=0.25$ and from 131 million paths at $x=0.35$ \cite{mthes}.
By construction, those data have a real part (drawn by solid line)
and an imaginary part (dashed line).
The imaginary part should vanish, hence its deviation from zero
gives an estimate of the statistical error in the real part.
The leading $x$ dependent term in the latter is expected to be proportional
to $x^2$. Since $0.35^2/0.25^2 \simeq 2$, we see from the plots that
the result of an extrapolation to $x=0$ would not be significantly
different from zero. 
One sees that our present calculation is about three orders of magnitude
more accurate than the Monte Carlo calculation. The $\sin^4\pi\nu$ term
that we find could not have been seen in that calculation, neither in the
one of ref.~\cite{num}, because of insufficient precision.

As originally noted in ref.~\cite{mo}, periodicity, supersymmetry
and analyticity imply that the third virial coefficient can be
represented as a rapidly converging Fourier series in only even powers of
$\sin\pi\nu$. The coefficient at the fourth power calculated here
is about 500 times smaller than the one at the second power.
A way to find it exactly would be to do fourth-order perturbation
theory in the manner of ref.~\cite{pert}. Our calculation is not precise
enough to show whether there are higher powers of $\sin\pi\nu$,
in fact the Fourier series might well be infinite.

The consistency of our present results with all previously known results
is another good check that they are reliable. In particular, that we
get a $\sin^2 \pi\nu$ term with the coefficient known from perturbation
theory is a nontrivial result. That the deviation from this $\sin^2 \pi\nu$
term has the form $\sin^4 \pi\nu$ is equally nontrivial. If we try to
fit it as $\sin^2 \pi\nu$, for example, we get $\chi^2/{\rm DOF} = 80$.

It is a remarkable result
that the conjecture (\ref{conj}) is valid within
about $0.2\%$, even though it is not exact.
This is likely to mean that there
exist some, as yet undiscovered, approximate analytic formulas
for the nonlinear energy levels, that yield
formula (\ref{conj}) and are valid within
a fraction of a per cent. It would be rather
interesting to understand what those formulas are.

\bigskip
We thank the Centre for Advanced Study for kind hospitality
and financial support.

\newpage

\begin{center}
\large \bf Figure captions
\end{center}

\begin{itemize}
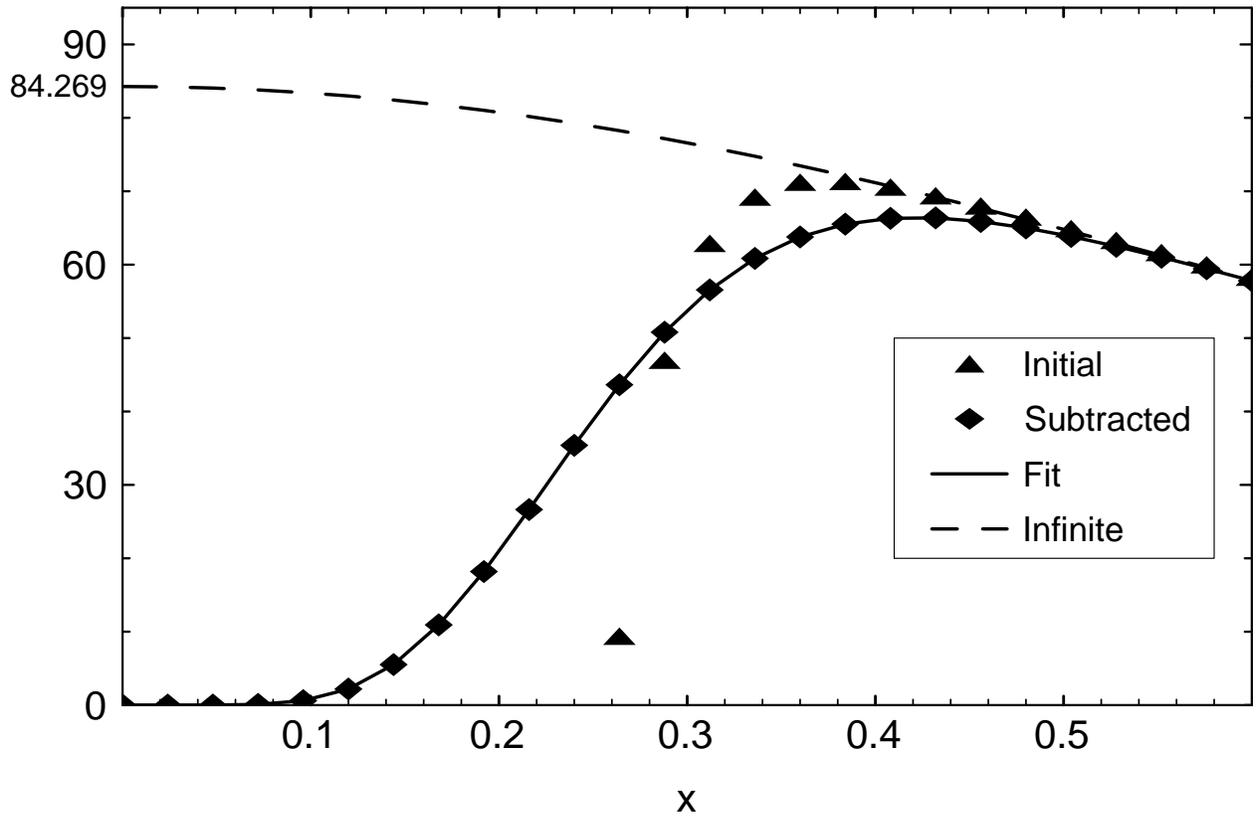
\item[Fig.~1.] Extracting $\Delta A_3(\nu)$ 
from the data for ${\cal E}=35$, at $\nu = 0.5$ (semions). \\
Triangles: $10^4\, \Delta A_3(\nu,x,{\cal E})$, eq.~(\ref{3}), versus $x$. \\
Diamonds: $10^4\, \Delta A_3^-(\nu,x,{\cal E})$, eq.~(\ref{minus}). \\
Solid line: the fit by eq.~(\ref{2fit}). \\
Dashed line: the ${\cal E}=\infty$ curve, eq.~(\ref{c2}). The $x=0$ point
of the latter is $\Delta A_3(\nu)$.

\bigskip

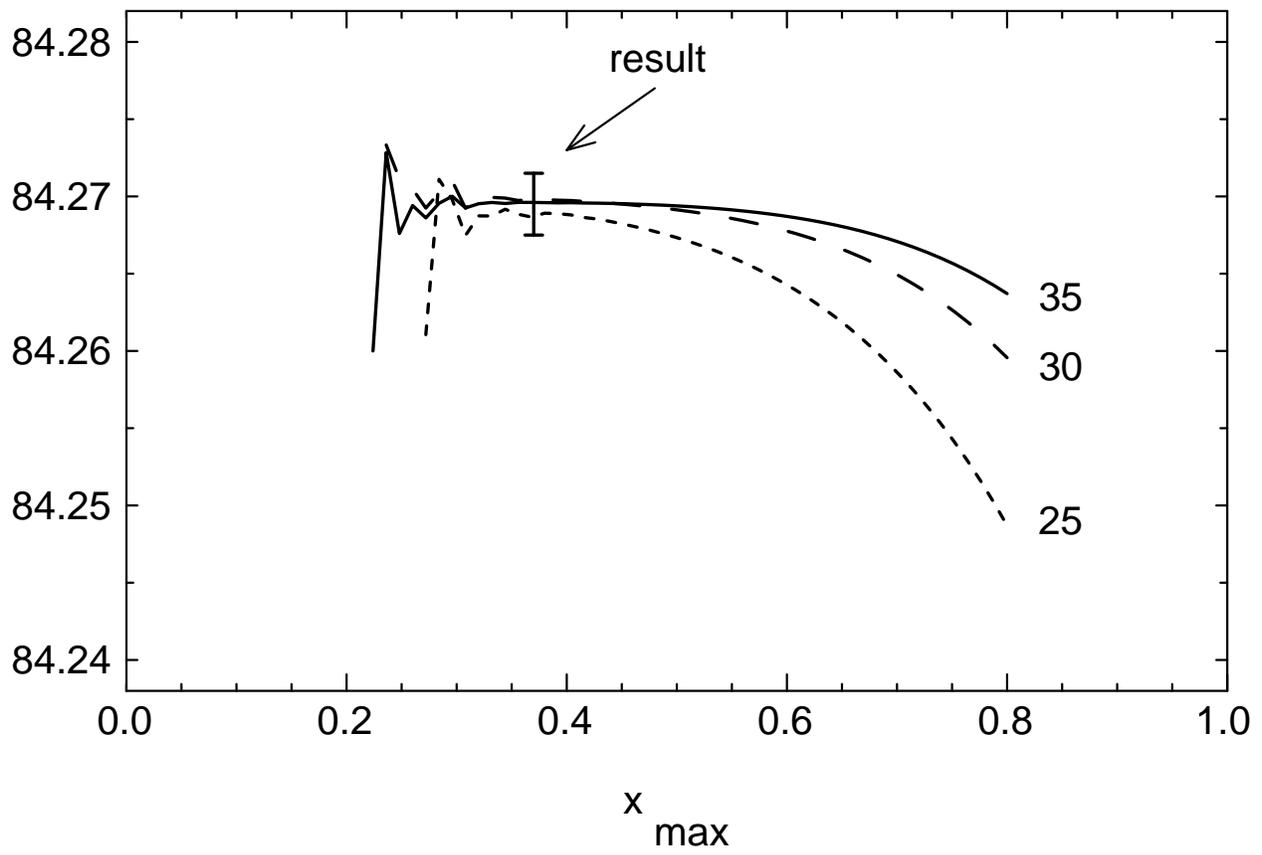
\item[Fig.~2.] The $x_{\rm max}$ dependence of the extracted
$10^4\, \Delta A_3(\nu)$ at $\nu = 0.5$. Curves for ${\cal E}=25,30$
and $35$ are plotted. Also shown is our final result with an error bar.
Note that $10^4/12\pi^2 = 84.43$ is well outside the plot.

\bigskip

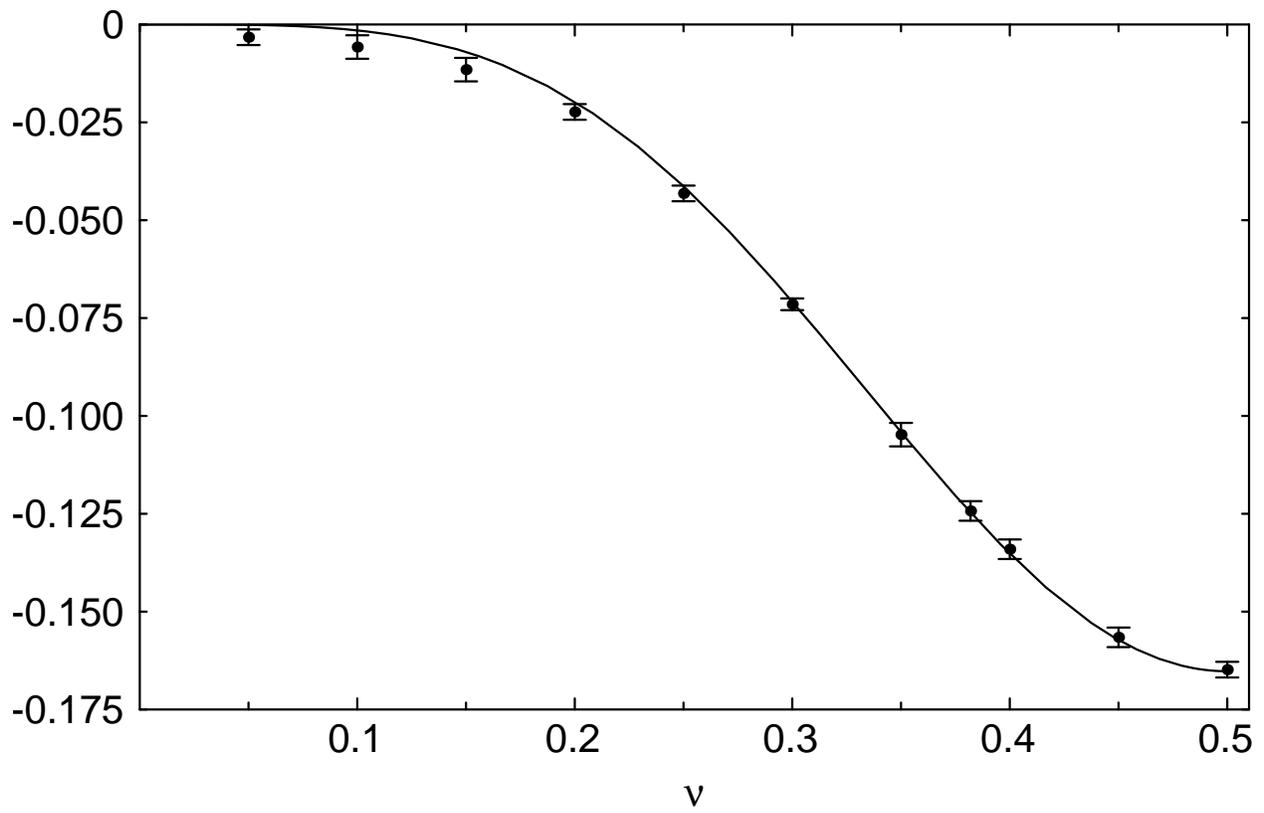
\item[Fig.~3.] The difference
$10^4\, [\Delta A_3(\nu)-(1/12\pi^2)\sin^2\pi\nu]$
as a function of $\nu$. The points with error bars are our
present results. The curve is a least square fit.

\bigskip

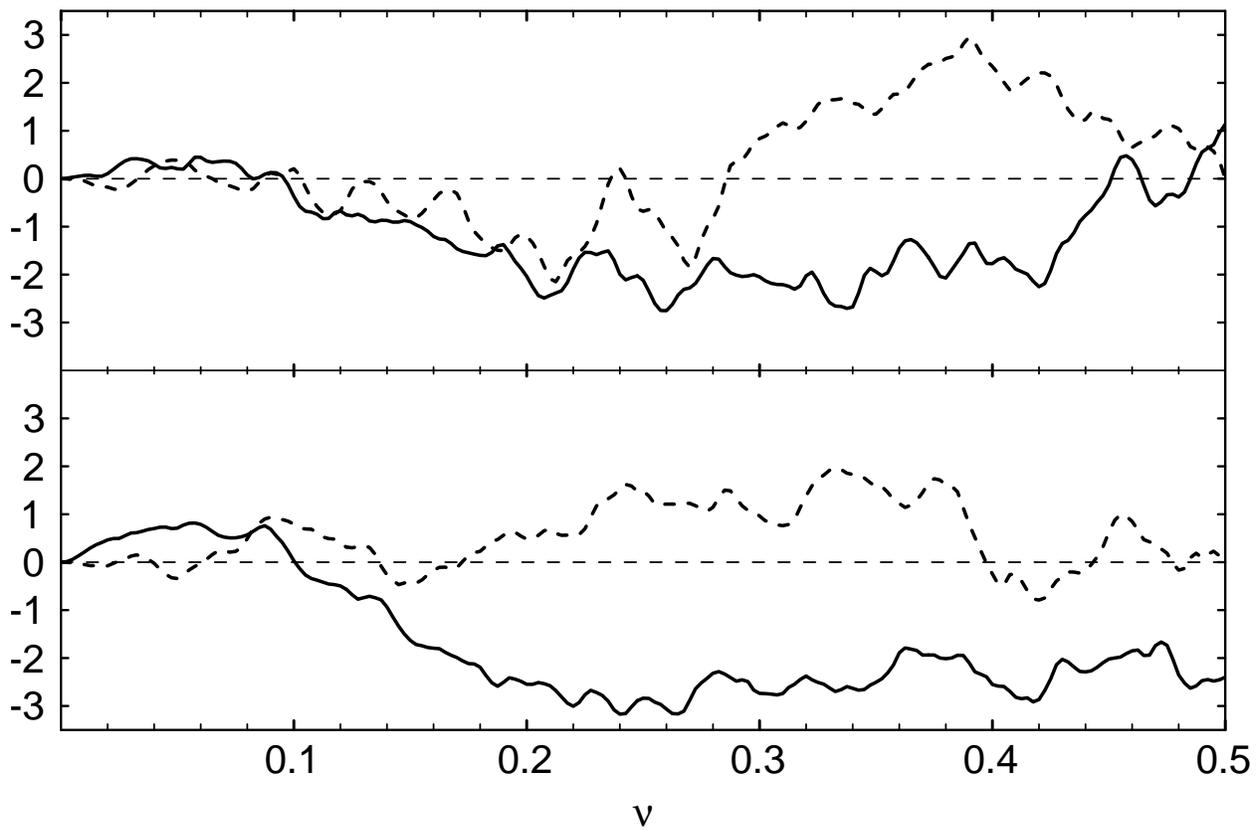
\item[Fig.~4.] The same quantity as in Fig.~3, from the Monte Carlo
simulation. \\
Solid lines: real part. \\
Dashed lines: imaginary part (illustrates the size of statistical errors). \\
The upper part is for $x=0.25$ (330 million paths). The lower part is for
$x=0.35$ (131 million paths).

\end{itemize}


\newpage
\begin{thebibliography}{99}
\bibitem{mmor1} S.~Mashkevich, J.~Myrheim, K.~Olaussen, R.~Rietman,
 Phys.\ Lett. B 348 (1995) 473.
\bibitem{2nd} D.~Arovas, J.R.~Schrieffer, F.~Wilczek, A.~Zee,
 Nucl.\ Phys. B 251 (1985) 117; \\
 J.S.~Dowker, J.\ Phys. A 18 (1985) 3521; \\
 A.~Comtet, Y.~Georgelin, S.~Ouvry, J.\ Phys. A 22 (1989) 3917.
\bibitem{mo} J.~Myrheim, K.~Olaussen, Phys.\ Lett. B 299 (1993) 267;
 305 (1993) 428(E).
\bibitem{mthes} J.~Myrheim, Dr.~philos.~thesis (1993); \\
 Notes for the Course on Geometric Phases, ICTP, Trieste (1993).
\bibitem{pert} A.~Dasni\`eres de Veigy, S.~Ouvry,
 Phys.\ Lett. B 291 (1992) 130; \\
 Nucl.\ Phys. B 388 (1992) 715.
\bibitem{ss} D.~Sen,
 Phys.\ Rev.\ Lett. 68 (1992) 2977; Phys.\ Rev. D 46 (1992) 1846.
\bibitem{numold} J.~Law, A.~Suzuki, R.K.~Bhaduri,
 Phys.\ Rev. A 46 (1992) 4693.
\bibitem{num} J.~Law, A.~Khare, R.K.~Bhaduri,
 A.~Suzuki, Phys.\ Rev. E 49 (1994) 1753.
\bibitem{tower} M.~Sporre, J.J.M.~Verbaarschot, I.~Zahed,
 Phys.\ Rev.\ Lett. 67 (1991) 1817; \\
 Nucl.~Phys. B 389 (1993) 645.
\bibitem{multi} S.~Mashkevich, Phys.\ Lett. B 295 (1992) 233;
 Phys.\ Rev. D 48 (1993) 5946.
\bibitem{ll} L.D.~Landau, E.M.~Lifshitz, Statistical Physics, Part 1,
 3rd edition (Pergamon Press, 1980), Sec.~33.
\end{thebibliography}
\end{document}